\documentclass{article}
\usepackage{spconf,amsmath,graphicx,hyperref}
\usepackage{subcaption}
\usepackage{times}
\usepackage{epsfig}
\usepackage{graphicx}
\usepackage{amsmath}
\usepackage{amssymb}
\usepackage{booktabs}
\usepackage{color}
\usepackage{multirow}
\usepackage{bbding}

\title{DITSINGER: SCALING SINGING VOICE SYNTHESIS WITH DIFFUSION TRANSFORMER AND IMPLICIT ALIGNMENT}
\name{%
\begin{tabular}{c}
\textit{Zongcai Du$^{*}$, Guilin Deng$^{*}$, Xiaofeng Guo$^{*}$, Xin Gao, Linke Li} \\
\textit{Kaichang Cheng, Fubo Han, Siyu Yang, Peng Liu, Pan Zhong, Qiang Fu}%
\thanks{$^*$Equal contribution.}
\end{tabular}%
}
\address{Migu Music, China Mobile Communications Corporation, China}

\begin{document}
\maketitle
%\footnotetext{$^*$Equal contribution.}
\begin{abstract}
Recent progress in diffusion-based Singing Voice Synthesis (SVS) demonstrates strong expressiveness but remains limited by data scarcity and model scalability. We introduce a two-stage pipeline: a compact seed set of human-sung recordings is constructed by pairing fixed melodies with diverse LLM-generated lyrics, and melody-specific models are trained to synthesize over 500 hours of high-quality Chinese singing data. Building on this corpus, we propose DiTSinger, a Diffusion Transformer with RoPE and qk-norm, systematically scaled in depth, width, and resolution for enhanced fidelity. Furthermore, we design an implicit alignment mechanism that obviates phoneme-level duration labels by constraining phoneme-to-acoustic attention within character-level spans, thereby improving robustness under noisy or uncertain alignments. Extensive experiments validate that our approach enables scalable, alignment-free, and high-fidelity SVS.
\end{abstract}

%-------------------------------------------------------------------------
\begin{figure*}[htbp]
	\centering
	\includegraphics[width=1.0\linewidth]{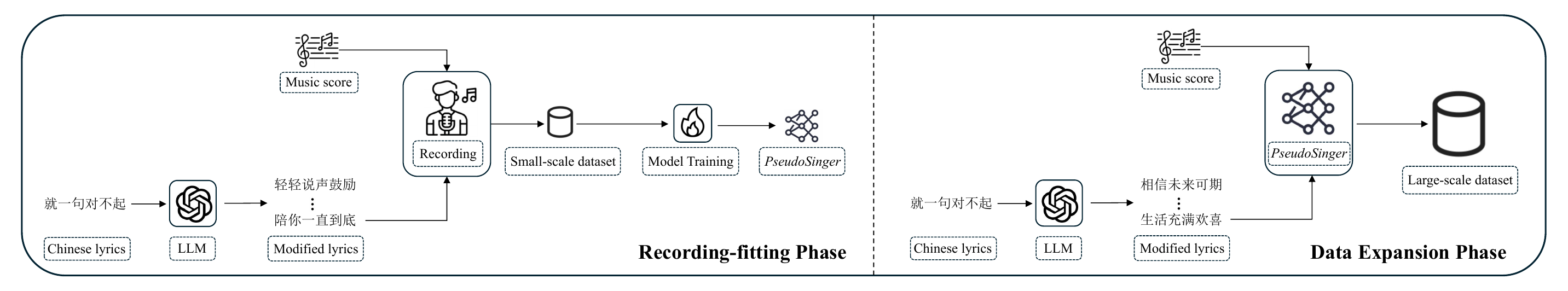} 
	\caption{Overview of the proposed two-stage data construction pipeline. The \textbf{Recording-fitting Phase} (left) collects high-quality vocal recordings without accompaniment from professional singers to train a melody-specific model, \textit{PseudoSinger}. The \textbf{Data Expansion Phase} (right) leverages the trained \textit{PseudoSinger} to synthesize large-scale singing data with diverse LLM-generated lyrics while keeping the melody fixed. This enables scalable dataset construction with improved phonetic consistency and melodic alignment.}
	\label{fig:datapipeline}
\end{figure*}
%-------------------------------------------------------------------------

\section{Introduction}
\label{sec:intro}
Singing voice synthesis (SVS) generates singing from lyrics and scores, requiring precise phoneme–pitch alignment and expressive modeling~\cite{survey3}. Early concatenative and statistical models~\cite{uc1, spm1} lacked naturalness, while neural approaches—from DNNs to GANs and non-autoregressive models~\cite{XiaoiceSing2, XiaoiceSing}—improved quality by reducing over-smoothing and exposure bias. Recent diffusion- and flow-based methods~\cite{DiffSinger, RMSSinger} offer finer timbre and technique control~\cite{StyleSinger, TechSinger, TCSinger}, and diverse datasets~\cite{M4Singer, GTSinger} support multiple vocal styles.

Despite recent advances, SVS faces two main challenges: unclear scaling effects on synthesis quality and limited methods for systematically expanding training data. We address this with a two-stage pipeline: fix a small set of melodies, use LLMs to generate diverse lyrics, pair with human recordings to train melody-specific models, and synthesize large-scale data with varied content, enhancing phonetic coverage and enabling controllable augmentation. To leverage the enlarged data and model scale, we design a Diffusion Transformer (DiT)~\cite{DiT} with rotary positional encoding (RoPE)~\cite{RoPE} and qk normalization~\cite{DeepSeekR1}.

The second challenge is robust phoneme-to-acoustic alignment. Prior methods rely on monotonic attention~\cite{DeepSinger} or duration prediction~\cite{DiffSinger, StyleSinger}, limiting flexibility and requiring post-processing. We propose an implicit cross-attention mechanism that constrains each phoneme’s attention to its character span, providing soft supervision and robustness under timing variability.

We present \textbf{DiTSinger}, a Diffusion Transformer-based SVS framework with strong scaling properties. Our main contributions are as follows:
\begin{itemize}
\item A scalable data pipeline combining LLM-generated lyrics with model-based audio synthesis to enhance phoneme diversity and generalization.
\item Introduction of DiT for SVS and analysis of scaling effects across data and model dimensions.
\item An implicit alignment mechanism linking phonemes to acoustic features at the character level, removing the need for duration annotations and improving timing robustness.
\end{itemize}

\section{Proposed Method}

\subsection{Preliminaries}
\noindent\textbf{Diffusion Models (DMs)}.\quad DDPM~\cite{DDPM} synthesize data by reversing a gradual noising process. The forward process corrupts a clean sample $\mathbf{x}_0$ via
\begin{equation}
q(\mathbf{x}_t \!\mid\! \mathbf{x}_{t-1}) = \mathcal{N}(\mathbf{x}_t; \sqrt{1 - \beta_t} \mathbf{x}_{t-1}, \beta_t \mathbf{I}),
\end{equation}
where $\mathcal{N}(\cdot)$ denotes the Gaussian distribution and $\beta_t$ is the noise schedule. The reverse process is modeled by a neural network $\boldsymbol{\epsilon}_\theta(\cdot)$ conditioned on $\mathbf{c}$ to predict the added noise:
\begin{equation}
\mathcal{L}_{\text{simple}} = \mathbb{E}_{\mathbf{x}_0, \boldsymbol{\epsilon}, t} \left[ \left\| \boldsymbol{\epsilon} - \boldsymbol{\epsilon}_\theta(\mathbf{x}_t, t, \mathbf{c}) \right\|_2^2 \right].
\end{equation}
Classifier-free guidance (CFG) improves fidelity and $w$ controls the guidance strength:
\begin{equation}
\boldsymbol{\epsilon}_{\text{guided}} = \boldsymbol{\epsilon}_\theta(\mathbf{x}_t) + w \cdot \left( \boldsymbol{\epsilon}_\theta(\mathbf{x}_t, \mathbf{c}) - \boldsymbol{\epsilon}_\theta(\mathbf{x}_t) \right).
\end{equation}
Latent Diffusion Models (LDMs) improve computational efficiency by encoding $\mathbf{x}_0$ into a latent representation $\mathbf{z}$~\cite{SoundStream,HifiGAN}:
\begin{equation}
\mathbf{z} = \text{Enc}(\mathbf{x}_0), \quad \hat{\mathbf{x}}_0 = \text{Dec}(\mathbf{z}).
\end{equation}

\subsection{Data Construction Pipeline}
Existing high-quality singing datasets are typically limited in scale, posing challenges for singing voice synthesis (SVS) models in capturing diverse pitch contours and phonetic variations. In particular, phoneme articulation and transitions can become unstable when models are exposed to unseen phonetic or linguistic content.

We observe that constraining training data to a small set of fixed melodies while varying only the lyrics and vocals reduces the complexity of learning melodic alignment and acoustic modeling. This strategy allows the model to internalize underlying melodic structures, facilitating more accurate and robust melody-conditioned synthesis across diverse lyrical inputs.

Motivated by this observation, we propose a two-stage data construction pipeline, illustrated in Figure~\ref{fig:datapipeline}, consisting of a \textbf{Recording-fitting Phase} and a \textbf{Data Expansion Phase}. In the Recording-fitting Phase, a small set of fixed melodies is paired with diverse lyric variants generated by a large language model (LLM). Professional singers record the corresponding clean vocals, resulting in a compact dataset used to train melody-specific SVS models, referred to as \textit{PseudoSinger}. In the subsequent Data Expansion Phase, each trained \textit{PseudoSinger} is leveraged to synthesize large-scale singing data. New lyrics are continually generated by the LLM and rendered into singing voices by \textit{PseudoSinger}, enabling scalable data generation while preserving melodic consistency.

To accelerate convergence and model phoneme transitions, we first train a \textit{base model} on the M4Singer~\cite{M4Singer} dataset. We then fine-tune 20 PseudoSinger models on disjoint groups of 500 melodies (50 rewrites each, 30h total) to synthesize ~500h of singing with consistent melodies and diverse lyrics, forming the largest publicly reported SVS dataset.

%-------------------------------------------------------------------------
\begin{figure*}[!t]
  \centering
  \includegraphics[width=\linewidth]{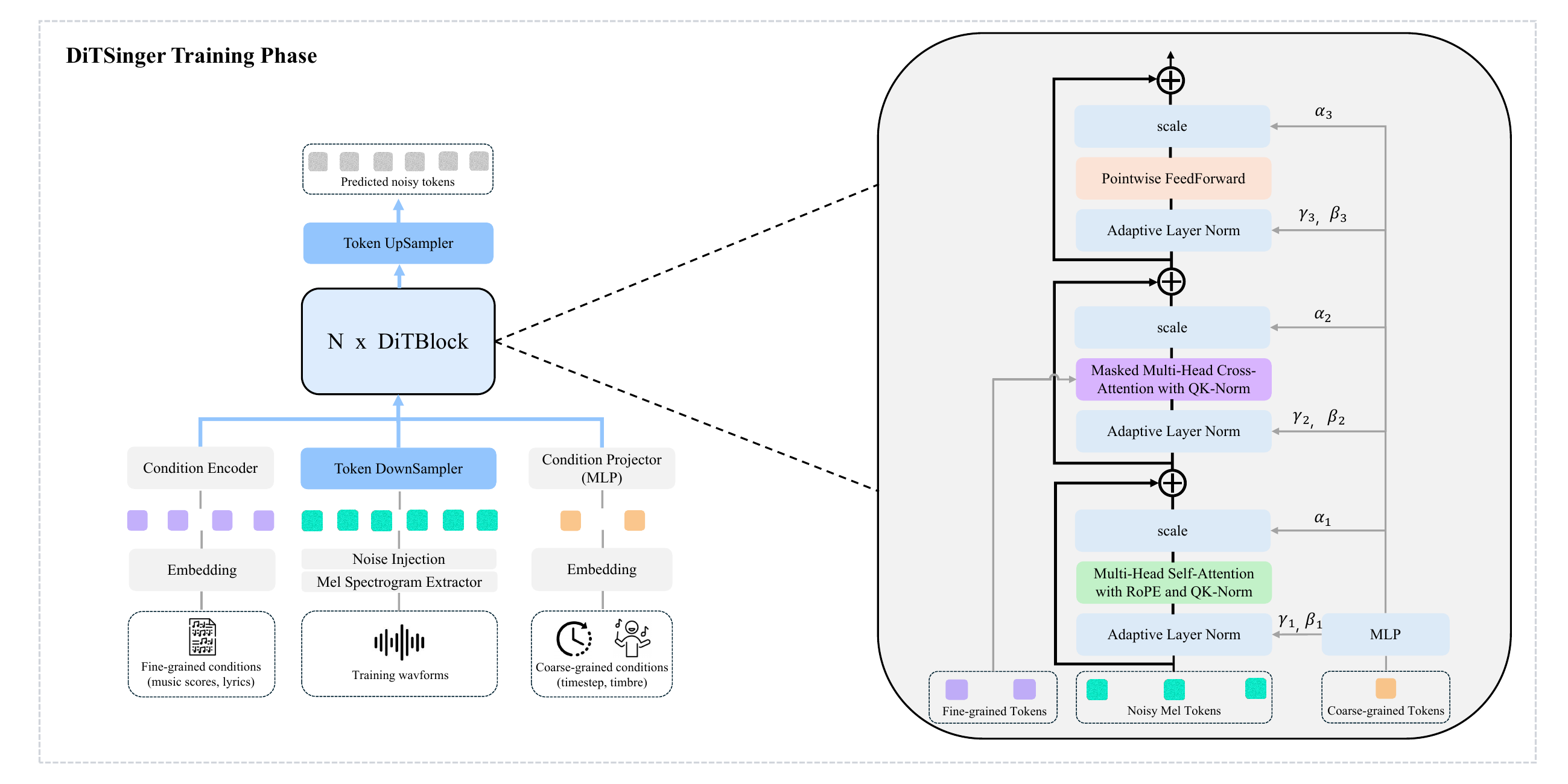}
  \caption{
    \textbf{DiTSinger Training Phase.}
    The model predicts the added noise $\boldsymbol{\epsilon}$ to the noisy mel-spectrogram tokens at each denoising step $t$, conditioned on both fine-grained (e.g., music scores, lyrics) and coarse-grained (e.g., timbre, timestep) inputs.
    Right: detailed structure of a single DiTBlock, which integrates Multi-Head Self-Attention with RoPE and QK-Norm, Multi-Head Cross-Attention with QK-Norm, and Adaptive Layer Normalization modulated by learnable parameters $\{\gamma_i, \beta_i\}$ and residual scaling factors $\{\alpha_i\}$.
  }
  \label{fig:architecture}
\end{figure*}
%-------------------------------------------------------------------------

\begin{figure*}[htbp]
    \centering
    \begin{minipage}[t]{0.45\textwidth}
        \centering
        \includegraphics[width=\linewidth]{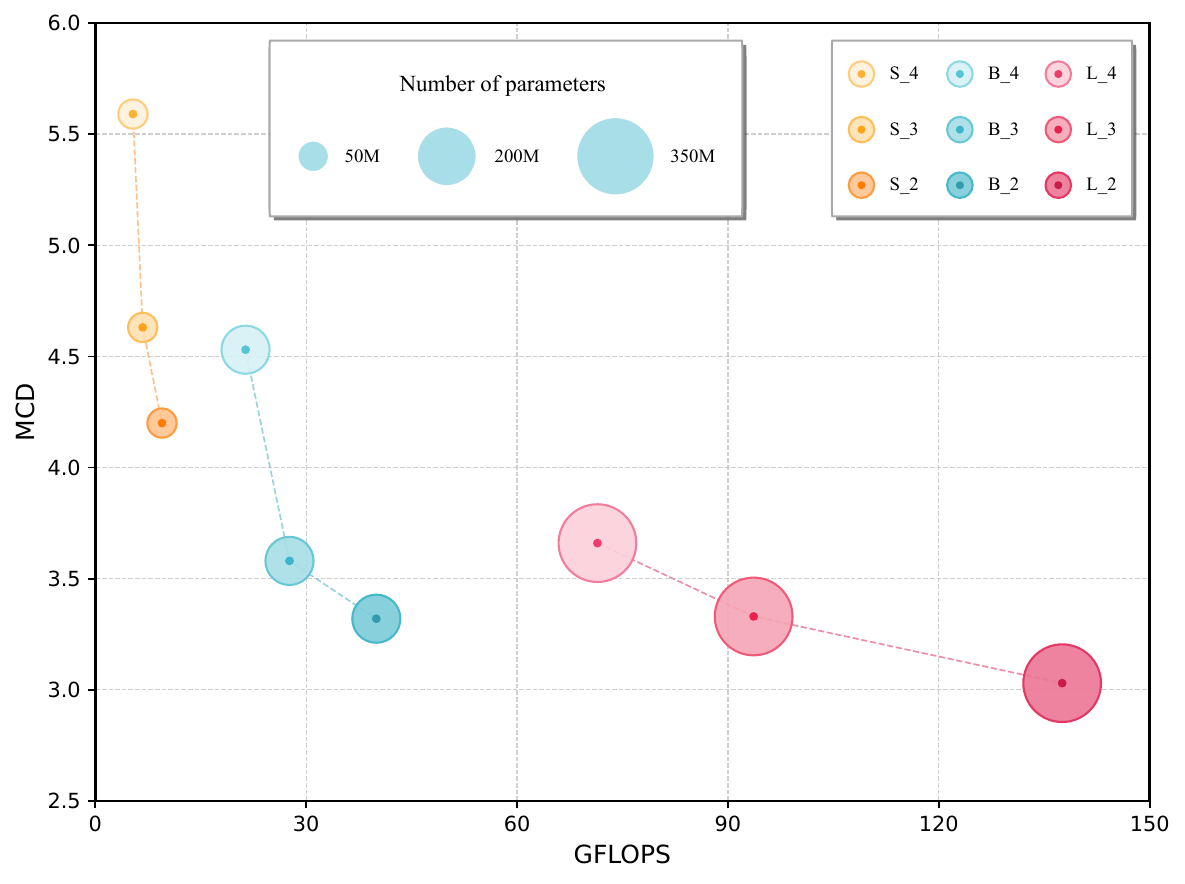}
        \par\smallskip
        (a)
    \end{minipage}\hfill
    \begin{minipage}[t]{0.45\textwidth}
        \centering
        \includegraphics[width=\linewidth]{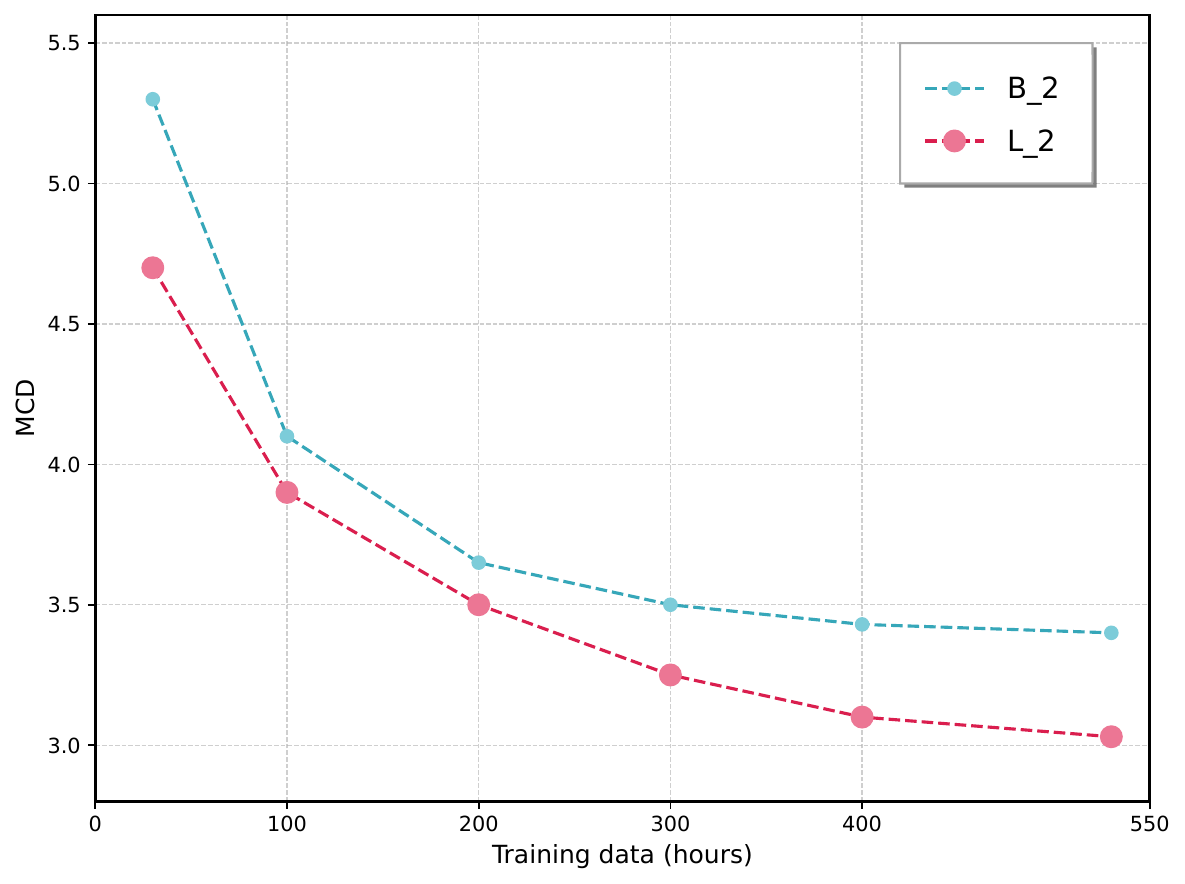}
        \par\smallskip
        (b)
    \end{minipage}

    \caption{Scaling results of DiTSinger. (a) Architectural scaling improves MCD. (b) Data scaling further boosts performance. S\_2 denotes a Small model with half resolution. GFLOPS measured on 5s audio.}
    \label{fig:model_scaling}
\end{figure*}

\subsection{Architecture}
Figure~\ref{fig:architecture} shows the training of \textbf{DiTSinger}, a transformer-based latent diffusion model that predicts noise $\boldsymbol{\epsilon}$ in the mel-spectrogram domain at each denoising step $t$.

\noindent\textbf{Conditioning inputs.}\quad DiTSinger uses hierarchical conditioning with fine- and coarse-grained information. Fine-grained inputs—pitch $\mathbf{p}$, phonemes $\mathbf{ph}$, word durations $\mathbf{w}$, and slur indicators $\mathbf{sl}$—are embedded, summed, and encoded via a Transformer-based condition encoder $\text{Enc}_{\text{cond}}$:
\begin{equation}
\mathbf{h}_{\text{local}} = \text{Enc}_{\text{cond}}(\mathbf{E}_{\text{p}}(\mathbf{p}) + \mathbf{E}_{\text{ph}}(\mathbf{ph}) + \mathbf{E}_{\text{w}}(\mathbf{w}) + \mathbf{E}_{\text{sl}}(\mathbf{sl})),
\end{equation}
where \(\mathbf{E}_{*}(\cdot)\) are learnable embeddings. Coarse-grained inputs, including speaker identity and diffusion timestep, are embedded via an MLP and injected through AdaLN~\cite{DiT}. Given the small number of speakers, timbre is represented with a learnable embedding table instead of a reference encoder.

\noindent\textbf{Tokenization and denoising.}\quad The waveform is converted to mel-spectrograms and tokenized into latents via a convolutional downsampler. Gaussian noise is added at each timestep $t$ to obtain $\mathbf{x}_t$ for diffusion training. The denoising network stacks $N$ DiTBlocks, each with three parallel branches: (1) Multi-Head Self-Attention (MHSA) with RoPE and QK-Norm, (2) Masked Multi-Head Cross-Attention (MHCA) incorporating fine-grained phoneme conditions, and (3) a pointwise FeedForward network. All branches use AdaLN conditioned on speaker embeddings, with residuals scaled by learnable parameters ${\alpha_1, \alpha_2, \alpha_3}$.

\noindent\textbf{Implicit Alignment Mechanism.}\quad We propose an \emph{Implicit Alignment Mechanism} to avoid costly phoneme-level duration labels. Each phoneme inherits its character’s temporal span, with known start time $t_{\text{start}}$ and duration $d_{\text{char}}$, extended backward by a tunable offset $\delta$:

\begin{equation}
\tilde{t}_{\text{start}} = t_{\text{start}} - \min(\delta, d_{\text{char}}, d_{\text{prev}}), \quad t_{\text{end}} = t_{\text{start}} + d_{\text{char}},
\end{equation}
where $d_{\text{prev}}$ denotes the duration of the preceding character. The resulting interval $[\tilde{t}_{\text{start}}, t_{\text{end}}]$ defines a valid interval used to construct an additive attention bias $M \in \mathbb{R}^{L_{\text{mel}} \times L_{\text{ph}}}$:
\begin{equation}
M_{i,j} =
\begin{cases}
0, & \text{if } t_i \in [\tilde{t}_{\text{start}}^{(j)}, t_{\text{end}}^{(j)}], \\
-\infty, & \text{otherwise}.
\end{cases}
\end{equation}

Let $Q \in \mathbb{R}^{L_{\text{mel}} \times d}$ be the query projected from mel tokens, and $K, V \in \mathbb{R}^{L_{\text{ph}} \times d}$ be the key and value projected from the fused local condition representation $\mathbf{h}_{\text{local}}$. The masked cross-attention is then computed as:
\begin{equation}
\mathrm{Attention}(Q, K, V) = \mathrm{softmax}\left(\frac{QK^\top}{\sqrt{d}} + M\right)V.
\end{equation}
This fixed mask is applied consistently during both training and inference. During training, it guides the model to learn soft and localized alignments under coarse timing constraints, supervised solely by the diffusion reconstruction loss. At inference time, it enforces the same temporal constraints to ensure stable and consistent attention patterns.

\section{Experiments}
\subsection{Settings}
\noindent\textbf{Datasets and evaluation metrics.}\quad We train on $\sim 530$h of singing from 40 professional vocalists, including data collected via our pipeline and the open-source M4Singer~\cite{M4Singer}, and evaluate on 50 segments from 10 songs excluded from training to test generalization to unseen melodies and lyrics. Synthesized singing is assessed with objective metrics MCD (DTW-aligned mel-cepstral coefficients from 24kHz, loudness-normalized), FFE (frames with voicing or pitch deviations $>$50 cents), F0RMSE (voiced frames), and a subjective MOS test (1–5 scale) with 95\% confidence intervals.

\noindent\textbf{Implementation and Baselines.}\quad We extract 80-bin mel-spectrograms from 24kHz audio (window 512, hop 128) with $\delta=1.0$. Training runs on 4 A100 GPUs for 100{,}000 iterations with per-GPU batch size 8 and 6-step gradient accumulation, using AdamW ($lr=0.001$) and 0.1 probability of dropping fine-grained conditions for classifier-free guidance. Inference uses DPM-Solver~\cite{DPM-Solver} with guidance scale 4.0, and training takes 3–7 days depending on model size. Baselines include Reference (human recording), Reference (vocoder, HiFi-GAN reconstruction), DiffSinger~\cite{DiffSinger} retrained on our dataset, and StyleSinger~\cite{StyleSinger} and TCSinger~\cite{TCSinger} conditioned on a reference clip from the same singer.

\subsection{Scalability of DiTSinger}
We investigate both model and data scaling. Model scaling is evaluated with Small (depth 4, width 384), Base(depth 8, width 576), and Large(depth 16, width 768) configurations using strided convolutions for resolution. Notably, S\_2 outperforms B\_4 despite lower complexity, underscoring the importance of resolution. Data scaling ranges from 30h to 530h. As shown in Figure~\ref{fig:model_scaling}, DiTSinger demonstrates strong scalability across both model size and dataset scale.

% \begin{table}[!htbp]
%   \centering
%   \caption{Effectiveness of PseudoSinger with different numbers of groups.}
%   \label{tab:pseudosinger}
  
%   \resizebox{0.5\textwidth}{!}{
%   \begin{tabular}{ccccc}
%     \toprule
%     \textbf{PseudoSinger \#} & \textbf{MOS} $\uparrow$ & \textbf{MCD} $\downarrow$ & \textbf{FFE} $\downarrow$ & \textbf{F0RMSE} $\downarrow$ \\
%     \midrule
%     1   & 3.62 $\pm$ 0.06 & 3.82 & 0.29 & 16.95 \\
%     10  & 3.88 $\pm$ 0.07 & 3.45 & 0.22 & 14.12 \\
%     20  & \textbf{4.05 $\pm$ 0.06} & \textbf{3.12} & \textbf{0.19} & \textbf{11.48} \\
%     30  & 4.02 $\pm$ 0.06 & 3.18 & 0.19 & 12.91 \\
%     40  & 3.98 $\pm$ 0.07 & 3.21 & 0.20 & 13.05 \\
%     50  & 3.81 $\pm$ 0.08 & 3.65 & 0.26 & 15.48 \\
%     \bottomrule
%   \end{tabular}
%   }
% \end{table}
\begin{table}[!htbp]
  \centering
  \caption{Effectiveness of PseudoSinger with different numbers of groups.}
  \label{tab:pseudosinger}
  
  \resizebox{0.5\textwidth}{!}{
  \begin{tabular}{ccccc}
    \toprule
    \textbf{PseudoSinger \#} & \textbf{MOS} $\uparrow$ & \textbf{MCD} $\downarrow$ & \textbf{FFE} $\downarrow$ & \textbf{F0RMSE} $\downarrow$ \\
    \midrule
    w/o base model & -- & -- & -- & -- \\
    1   & 3.62 $\pm$ 0.06 & 3.82 & 0.29 & 16.95 \\
    10  & 3.88 $\pm$ 0.07 & 3.45 & 0.22 & 14.12 \\
    20  & \textbf{4.05 $\pm$ 0.06} & \textbf{3.12} & \textbf{0.19} & \textbf{11.48} \\
    30  & 4.02 $\pm$ 0.06 & 3.18 & 0.19 & 12.91 \\
    40  & 3.98 $\pm$ 0.07 & 3.21 & 0.20 & 13.05 \\
    50  & 3.81 $\pm$ 0.08 & 3.65 & 0.26 & 15.48 \\
    \bottomrule
  \end{tabular}
  }
\end{table}

\subsection{Effectiveness of PseudoSinger}
We evaluate PseudoSinger by varying the number of groups from 1 to 50 (Table~\ref{tab:pseudosinger}), measuring metrics on training-set MIDI with out-of-set lyrics to assess melody fitting and generalization. With one group (base model), melodic contours are captured but articulation is unstable. Performance improves with more groups, peaking at 20, then saturates; at 50 groups, where each PseudoSinger has fewer MIDIs, generalization worsens. These results suggest a moderate number of groups balances specialization and generalization.

\subsection{Comparison with State-of-the-Art Methods}
We compare DiTSinger with representative state-of-the-art SVS models, including DiffSinger~\cite{DiffSinger}, StyleSinger~\cite{StyleSinger}, and TCSinger~\cite{TCSinger}. As shown in Table~\ref{tab:metrics}, DiTSinger\_L\_2 achieves the best overall performance, yielding the highest MOS and consistently lower MCD, FFE, and F0RMSE. Notably, DiTSinger\_L\_2 surpasses DiffSinger (retrained on our data) by 0.22 MOS and significantly reduces F0 errors, highlighting the effectiveness of our implicit alignment framework.
\begin{table}[!htbp]
  \centering
  \caption{Comparison of DiTSinger variants with baselines on MOS, MCD (dB), FFE, and F0RMSE (Hz). DiffSinger~\cite{DiffSinger} is retrained on our data.}
  \label{tab:metrics}
  
  \resizebox{0.5\textwidth}{!}{
  \begin{tabular}{lcccc}
    \toprule
    \textbf{Method} & \textbf{MOS} $\uparrow$ & \textbf{MCD} $\downarrow$ & \textbf{FFE} $\downarrow$ & \textbf{F0RMSE} $\downarrow$ \\
    \midrule
    Reference & 4.35 $\pm$ 0.04 & -- & -- & -- \\
    Reference (vocoder) & 4.12 $\pm$ 0.06 & 1.45 & 0.06 & 3.60 \\
    \midrule
    DiffSinger~\cite{DiffSinger} & 3.80 $\pm$ 0.06 & 3.54 & 0.24 & 14.15 \\
    StyleSinger~\cite{StyleSinger} & 3.62 $\pm$ 0.08 & 3.78 & 0.28 & 16.72 \\
    TCSinger~\cite{TCSinger} & 3.89 $\pm$ 0.06 & 3.51 & 0.22 & 13.83 \\
    \midrule
    DiTSinger\_S\_2  & 3.47 $\pm$ 0.09 & 4.12 & 0.32 & 17.83 \\
    DiTSinger\_B\_2  & 3.95 $\pm$ 0.05 & 3.38 & 0.18 & 13.25 \\
    \textbf{DiTSinger\_L\_2}  & \textbf{4.02 $\pm$ 0.06} & \textbf{3.03} & \textbf{0.15} & \textbf{11.18} \\
    \bottomrule
  \end{tabular}
  }
\end{table}

\section{Limitations and Future Work}
Although we propose a scalable data augmentation pipeline and architecture, experiments are limited to Chinese datasets, and the model ignores factors like singing techniques. Future work will expand datasets and incorporate conditions such as reference timbre and singing style to improve multilingual and multi-scenario adaptability.

\vfill\pagebreak


\bibliographystyle{IEEEbib}
\bibliography{refs}

\end{document}